\documentstyle{article}

\input pz.sty
\input epsf.sty

\begin{document}

\PZhead{4}{27}{2007}{3 October}

\PZtitle{Photometric observations of Supernova 2002\lowercase{hh}}

\PZauth{D. Yu. Tsvetkov$^1$, M. M. Muminov$^2$, O. A. Burkhanov$^2$, 
B. B. Kahharov$^2$}
\PZinst{Sternberg Astronomical Institute, University Ave.13,
119992 Moscow, Russia; e-mail: tsvetkov@sai.msu.su}
\PZinst{Ulugh Bek Astronomical Institute of Uzbekistan Academy
of Sciences}

\begin{abstract}
CCD $VRI$ photometry is presented for 
SN 2002hh from 14 days after the outburst till day 347. 
SN 2002hh appears to be normal type IIP supernova
regarding both luminosity and the shape of the light curve, which
is similar to SN 1999gi.  
\end{abstract}

\begintext

SN 2002hh was discovered on 2002 October 31.1 UT during the course of
the Lick Observatory Supernova Search (Li, 2002). SN is located
at $\alpha  = 20\hr34\mm44\sec.29, \delta = +60\deg07\arcm19\arcs.0$
(2000.0), which is $60\arcs.9$ west and $114\arcs.1$ south of the 
nucleus of Scd galaxy NGC 6946, which has produced 7 other SNe.
Spectra taken by Filippenko et al. (2002) on 2002 November 2
revealed it to be a very young, highly reddened type II SN.
Broad, low-contrast H$\alpha$ emission and absorption lines as well
as strong, narrow interstellar Na I D absorption were present.
The continuum was nearly featureless and very red.
SN 2002hh was also detected as a source of radio and X-ray
emission (Stockdale et al., 2002; Pooley and Lewin, 2002).  

A detailed study of optical and infrared photometric and spectroscopic
evolution for this object was presented by Pozzo et al. (2006). 
They concluded that
SN 2002hh was a SN IIP (plateau), with early light curve similar to
SN IIP 1999em, and that radioactive tails are well matched for 
these two SNe and SN 1987A. They adopted two-component model for
extinction with total $A_V=5.2$ mag.

\medskip 

We observed SN 2002hh from 2002 November 13 until 2003 October 12
with different telescopes and detectors: 60-cm reflector of Crimean 
Observatory of Sternberg Astronomical Institute (C60) equipped with
SBIG ST-7 CCD camera; 70-cm reflector in Moscow (M70) with Meade Pictor416XT
camera (a) or Apogee AP47 camera (b); 1.5-m reflector of Maidanak
Observatory (Md150) with SITe 2000x800 LN cooled CCD camera. 

The image of SN 2002hh obtained at Md150 on 2003 August 10 in the
$I$ band is shown in Fig. 1,
where the local standard stars are marked. The magnitudes of these
stars were measured on 11 photometric nights mostly in 2004-2005, when
observations of SN 2004et in the same galaxy were carried out, they are
reported in Table 1. $VRI$ magnitudes of these stars were derived also by
Pozzo et al. (2006), and the mean differences 
between two data sets
and their dispersions are:
$\overline{\Delta V}= 0.015; \sigma_{\Delta V}=0.01; 
\overline{\Delta R}=-0.008; 
\sigma_{\Delta R}=0.01; \overline{\Delta I}=0.053; 
\sigma_{\Delta I}=0.027$.
We may conclude that the agreement is good.
Photometric measurements of SN were made relative to 
local standard stars using PSF-fitting with IRAF\PZfm
\PZfoot{IRAF is distributed by the National Optical Astronomy Observatory,
which is operated by AURA under cooperative agreement with the
National Science Foundation}
DAOPHOT package. On the nights with bad seeing the images of SN and 
nearby bright star overlayed, and it was necessary to subtract the
image of this star using task SUBSTAR in DAOPHOT. 

The color terms for transformation of instrumental magnitudes $vri$
to standard $VRcIc$ were derived from observations of standard fields for 
C60 and M70 and by photometry of local standards for Md150. The resulting
color terms are $K_v=-0.007; K_r=-0.48; K_i=-0.27$ for C60;
$K_r=-0.35; K_i=-0.31$ for M70a; $K_r=-0.46; K_i=-0.37$ for M70b;
$K_r=0.12; K_i=0.0$ for Md150.   

The photometry of SN 2002hh is presented in Table 2, and the light
curves are shown in Fig. 2, where the data from Pozzo et al. (2006)
and magnitude estimates at discovery 
and prediscovery upper limit from Li (2002) are also plotted.
At the plateau stage the agreement between the 
two data sets is quite good, although our $R$ and $I$ filters
at C60 and M70 
do not match the standard system well. Only in the $I$ band there
is evidence for some systematic difference, our magnitudes being 
about 0.15 mag brighter.
 
We obtained images on two dates (2003 March 3 and March 26)
which are in the gap of Pozzo et al. (2006) data. At the exponential
decline stage our $I$ magnitudes from Md150 are in very good agreement
with the results of Pozzo et al. (2006), while in the $R$ band our 
magnitudes are systematically brighter by about 0.18 mag. This 
difference is likely due to different response curves of the equipment
applied to the object with very red color and emission-dominated spectrum. 
This is also the reason for our magnitudes from M70 to be brighter than
from Md150; the later should be given greater weight as their errors
are smaller and the color system closer to standard.

We estimated the rate of decline at the exponential tail
by fitting straight line to the data from Md150:
$0.0073\pm 0.0003$ mag day$^{-1}$ in $R$ and
$0.0094\pm 0.0004$ mag day$^{-1}$ in $I$. The result for $I$ band is
close to that by Pozzo et al. (2006), but in $R$ they found
significantly larger rate: 0.011 mag day$^{-1}$. We suppose that
the difference is due to larger errors of magnitudes by Pozzo et al.
(2006) at late stage. We can also estimate the drop of brightness
from the plateau to the onset of exponential tail: 1.4 mag in 
$R$ and 1.6 mag in $I$. 

We found that the light curve of SN 1999em was not a good match to 
the data for SN 2002hh, as the brightness decline  between the plateau
and the onset of exponential tail was significantly smaller for
SN 2002hh. Among the well studied SN IIP with normal luminosity
SN 1999gi was probably the best match, although 
considerable differences can be seen on Fig. 2, where we plotted
the light curves of SN 1999gi according to Leonard et al. (2002)
and our own unpublished data. While in the $I$ band the match is
very good, in $R$ and $V$ the magnitude difference between plateau and 
the onset of final decline is greater for SN 1999gi.   

Taking the extinction from Pozzo et al. (2006) and a host galaxy distance
of 5.9 Mpc (Karachentsev et al. 2000), we obtain absolute magnitude
at maximum light $M_V=-16.7$, close to the mean value for SN IIP
(Richardson et al., 2002). 

\medskip

We conclude that SN 2002hh is a normal type IIP supernova, with 
the light curve similar to SN 1999gi, especially in the $I$ band.
Our photometry confirms the results of Pozzo et al. (2006), but we
obtain slightly different magnitudes and rate of decline in the $R$
band at the epoch 200-340 days past explosion. 
  
\medskip

The work of D.Yu.Tsvetkov was partly supported by the grant
05-02-17480 of the Russian Foundation for Basic Research. 

\newpage 

\references

Filippenko, A.V., Foley, R.J., Swift, B., 2002,
{\it IAU Circ.}, No. 8007

Karachentsev, I.D., Sharina, M.E., Huchtmeier, W.K., 2000,
{\it Astron.\& Astrophys.}, {\bf 362}, 544

Leonard, D.C., Filippenko, A.V., Li, W., et al., 2002,
 {\it Astron. J.}, {\bf 124}, 2490 

Li, W., 2002, {\it IAU Circ.}, No. 8005

Stockdale, C.J., Sramek, R.A., Rupen, M.,
Weiler, K.W., Van Dyk ,S.D., Panagia, N., Pooley, D., Lewin, W.,
Myers, S., Taylor, G., 2002, {\it IAU Circ.}, No. 8018

Pooley, D., Lewin, W.H.G., 2002, {\it IAU Circ.}, No. 8024 

Pozzo, M., Meikle, W.P.S., Rayner, J.T., Joseph, R.D.,  
Filippenko, A.V., Foley, R.J., Li, W., Mattila, S., Sollerman, J.,
2006, {\it MNRAS}, {\bf 368}, 1169

Richardson, D., Branch, D., Casebeer, D., et al., 2002, {\it Astron. J.}
{\bf 123}, 745

\endreferences

\PZfig{11cm}{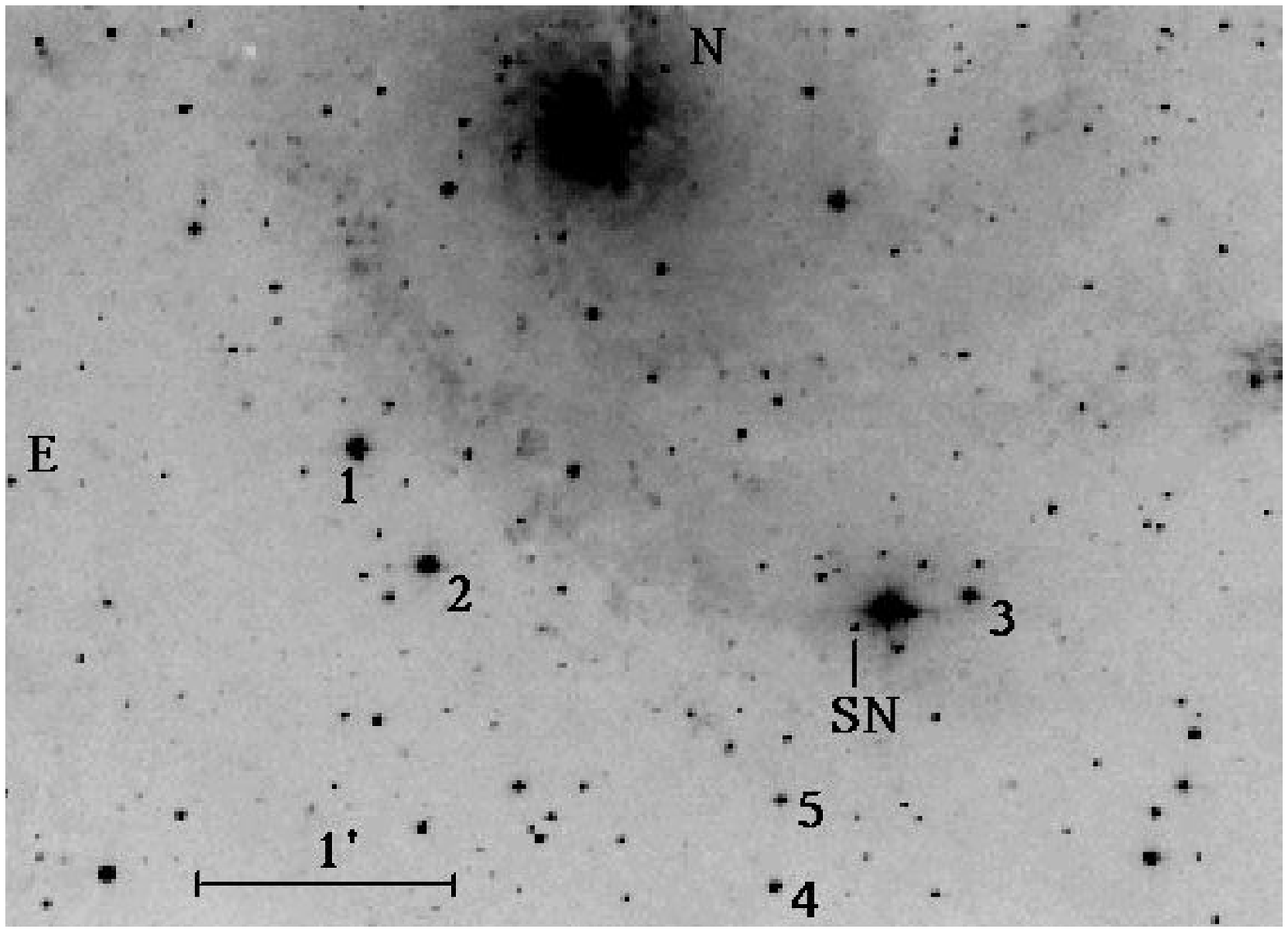}{SN 2002hh and local standard stars}

\newpage

\PZfig{13cm}{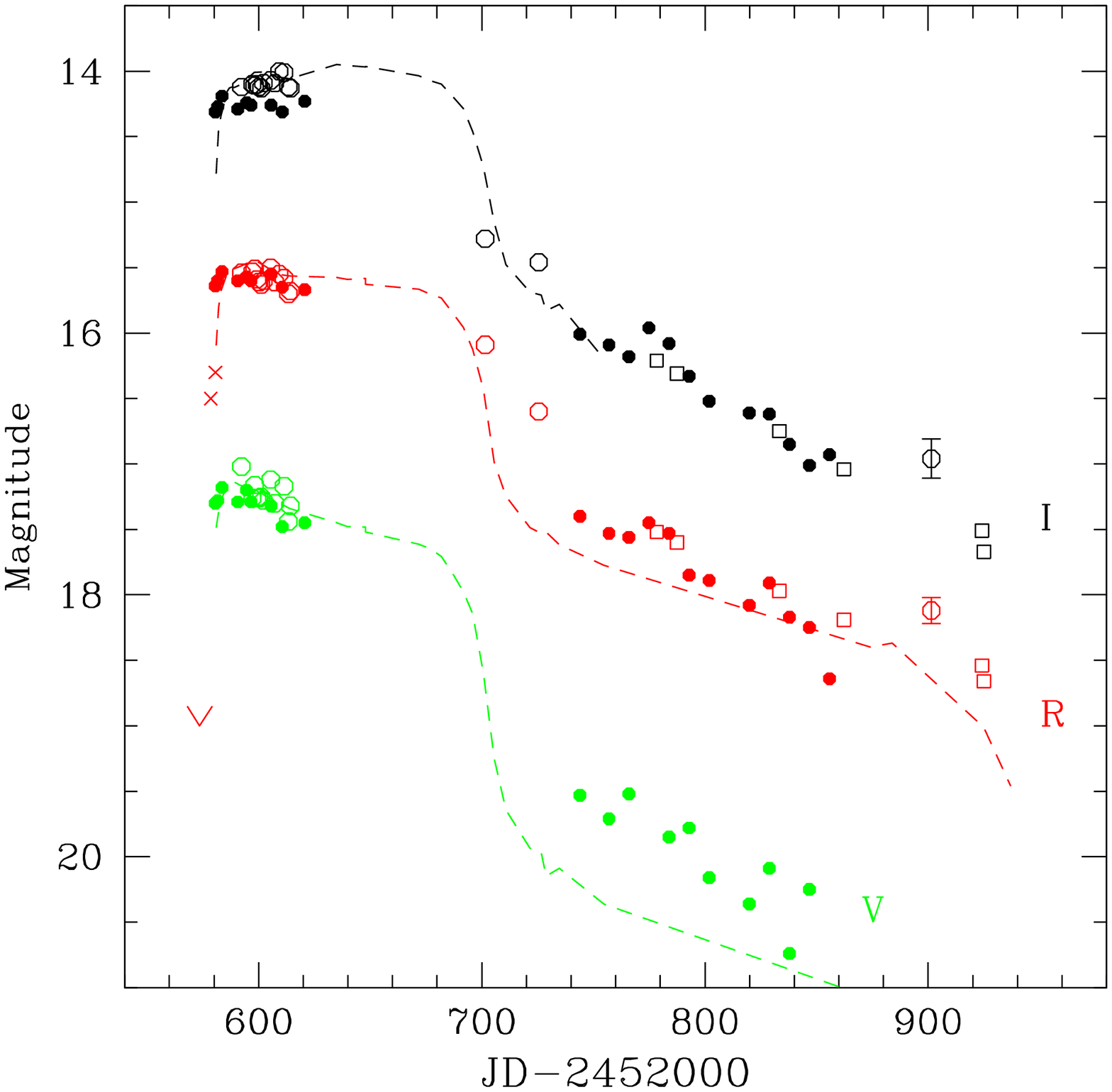} {$VRI$ light curves of SN 2002hh.
Circles show our data from C60 and M70, squares are for our data
from Md150, dots show photometry by Pozzo et al. (2006), crosses and
'v' mark are for discovery estimates and prediscovery limit by 
Li (2002). Error bars for our magnitudes are plotted only when they
exceed the size of a point. The dashed lines are the light curves of
SN 1999gi} 

\newpage

\begin{table}
\centering
\caption{Magnitudes of local standard stars}\vskip2mm
\begin{tabular}{ccccccccccc}
\hline
Star & $U$ & $\sigma_U$ & $B$ & $\sigma_B$ & $V$ & $\sigma_V$ & 
$R$ & $\sigma_R$ & $I$ & $\sigma_I$\\
\hline
1 & 14.60 & 0.08 & 14.30 & 0.01 & 13.56 & 0.01 & 13.14 & 0.03 & 12.75 & 0.02 \\
2 & 14.78 & 0.03 & 14.51 & 0.02 & 13.77 & 0.01 & 13.35 & 0.03 & 12.96 & 0.02 \\
3 & 15.92 & 0.03 & 15.59 & 0.03 & 14.76 & 0.01 & 14.22 & 0.04 & 13.85 & 0.03 \\
4 &       &      & 16.62 & 0.05 & 15.86 & 0.01 & 15.35 & 0.04 & 15.00 & 0.03 \\
5 &       &      & 17.58 & 0.05 & 16.46 & 0.06 & 15.74 & 0.04 & 15.17 & 0.02 \\  
\hline
\end{tabular}
\end{table}

\begin{table}
\centering   
\caption{Photometry of SN 2002hh}\vskip2mm
\begin{tabular}{cccccccl}
\hline
JD 2452000+ & $V$ & $\sigma_V$ & $R$ & $\sigma_R$ & $I$ & $\sigma_I$ & Tel.\\
\hline
592.31 &  17.02 & 0.07 & 15.54 & 0.03 & 14.12 & 0.03 & C60 \\
597.21 &  17.26 & 0.06 & 15.53 & 0.07 & 14.10 & 0.04 & C60 \\
598.23 &  17.16 & 0.05 & 15.51 & 0.04 & 14.11 & 0.04 & C60 \\
599.24 &  17.26 & 0.05 & 15.59 & 0.04 & 14.10 & 0.04 & C60 \\
600.20 &  17.26 & 0.05 & 15.61 & 0.05 & 14.12 & 0.04 & C60 \\
601.17 &  17.25 & 0.08 & 15.63 & 0.05 & 14.13 & 0.07 & C60 \\
602.18 &  17.28 & 0.05 & 15.60 & 0.03 & 14.09 & 0.03 & C60 \\
605.38 &  17.12 & 0.09 & 15.50 & 0.04 & 14.07 & 0.04 & C60 \\
607.26 &  17.30 & 0.08 & 15.61 & 0.05 & 14.09 & 0.05 & C60 \\
609.32 &        &      & 15.55 & 0.04 & 14.00 & 0.05 & C60 \\
611.38 &  17.17 & 0.07 & 15.58 & 0.04 & 14.01 & 0.07 & C60 \\
613.39 &  17.44 & 0.05 & 15.70 & 0.03 & 14.12 & 0.03 & C60 \\
614.27 &  17.32 & 0.07 & 15.68 & 0.05 & 14.13 & 0.04 & C60 \\
701.50 &        &      & 16.09 & 0.05 & 15.28 & 0.04 & M70a \\
725.54 &        &      & 16.60 & 0.07 & 15.46 & 0.04 & M70a \\
778.44 &        &      & 17.52 & 0.03 & 16.21 & 0.02 & Md150 \\
787.46 &        &      & 17.60 & 0.03 & 16.31 & 0.02 & Md150 \\
833.38 &        &      & 17.97 & 0.04 & 16.75 & 0.02 & Md150 \\
862.27 &        &      & 18.19 & 0.03 & 17.04 & 0.02 & Md150 \\
901.44 &        &      & 18.12 & 0.11 & 16.96 & 0.16 & M70b \\
924.13 &        &      & 18.54 & 0.03 & 17.51 & 0.02 & Md150\\
925.10 &        &      & 18.66 & 0.03 & 17.67 & 0.02 & Md150\\
\hline
\end{tabular}
\end{table}
\end{document}